\documentclass[twocolumn]{aastex631}

\usepackage{amsmath, mathtools, amsthm, amssymb}
\usepackage{enumitem}
\usepackage[normalem]{ulem}
\usepackage[toc,page]{appendix}
\usepackage{soul}
\usepackage{graphicx}
\graphicspath{ {./} }
\usepackage{xcolor}

\usepackage{booktabs}

\usepackage{hyperref}
\hypersetup{colorlinks=true, allcolors=cyan}

\def\HAGN{\textsc{Horizon-AGN}} 
\def\NH{\textsc{NewHorizon}}
\def\SAMI{\textsc{SAMI}}
\def\CALIFA{\textsc{CALIFA}}
\def\IRAF{\textsc{IRAF}}
\def\GALFIT{\textsc{GALFIT}}
\def\PROFIT{\textsc{PROFIT}}
\def\RAMSES{\textsc{RAMSES}}
\def\SKIRT{\textsc{SKIRT}}

\def\sersic{S\'{e}rsic}

\def\dttk{$[D/T]_{\rm kin}$}
\def\dttp{$[D/T]_{\rm phot}$}

\def\vsig{$V/\sigma$}
\def\spinp{$\lambda_{\rm R}$}
\def\kppar{${\kappa}_{\rm rot}$}

\def\dx{$\Delta x$}
\def\reff{$R_{\rm e}$}
\def\rft{$R_{\rm 50}$}
\def\rnt{$R_{\rm 90}$}

\def\msol{$\rm M_{\rm \odot}$}

\def\cpar{$J_{\rm z}\,/\,J_{\rm cir}(e)$}
\def\ppar{$J_{\rm p}\,/\,J_{\rm cir}(e)$}
\def\epar{$e\,/\,|e_{\rm max}|$}

\definecolor{Green}{rgb}{0.1,0.7,0.2}

\newcommand\edel{\bgroup\markoverwith
{\textcolor{red}{\rule[0.5ex]{2pt}{0.8pt}}}\ULon}

\accepted{by ApJ. April 13, 2023}

\shorttitle{Translators of galaxy morphology indicators}

\shortauthors{Jang, JK et al.}

\begin{document}

\title{Translators of galaxy morphology indicators between observation and simulation}

\author{J. K. Jang}
\affil{Department of Astronomy and Yonsei University Observatory, Yonsei University, Seoul 03722, Korea}

\author{Sukyoung K. Yi}
\affil{Department of Astronomy and Yonsei University Observatory, Yonsei University, Seoul 03722, Korea}
\email{yi@yonsei.ac.kr}

\author{Yohan Dubois}
\affiliation{Institut d’Astrophysique de Paris, Sorbonne Université, CNRS, UMR 7095, 98 bis bd Arago, 75014 Paris, France }

\author{Jinsu Rhee}
\affil{Department of Astronomy and Yonsei University Observatory, Yonsei University, Seoul 03722, Korea}

\author{Christophe Pichon}
\affiliation{Institut d’Astrophysique de Paris, Sorbonne Université, CNRS, UMR 7095, 98 bis bd Arago, 75014 Paris, France }

\author{Taysun Kimm}
\affil{Department of Astronomy and Yonsei University Observatory, Yonsei University, Seoul 03722, Korea}

\author{Julien Devriendt}
\affiliation{Dept of Physics, University of Oxford, Keble Road, Oxford OX1 3RH, UK}

\author{Marta Volonteri}
\affiliation{Institut d’Astrophysique de Paris, Sorbonne Université, CNRS, UMR 7095, 98 bis bd Arago, 75014 Paris, France }

\author{Sugata Kaviraj}
\affiliation{University of Hertfordshire, Hatfield, Hertfordshire, United Kingdom}

\author{Sebastien Peirani}
\affiliation{Université Côte d’Azur, Observatoire de la Côte d’Azur, CNRS, Laboratoire Lagrange, Nice, France}

\author{Sree Oh}
\affil{Department of Astronomy and Yonsei University Observatory, Yonsei University, Seoul 03722, Korea}
\affiliation{Research School of Astronomy and Astrophysics, Australian National University, Canberra, ACT 2611, Australia}
\affiliation{ARC Centre of Excellence for All Sky Astrophysics in 3 Dimensions (ASTRO 3D), Australia}

\author{Scott Croom}
\affiliation{ARC Centre of Excellence for All Sky Astrophysics in 3 Dimensions (ASTRO 3D), Australia}
\affiliation{Sydney Institute for Astronomy (SIfA), School of Physics, The University of Sydney, NSW 2006, Australia}




\begin{abstract}
Based on the recent advancements in the numerical simulations of galaxy formation, we anticipate the achievement of realistic models of galaxies in the near future. 
Morphology is the most basic and fundamental property of galaxies, yet observations and simulations still use different methods to determine galaxy morphology, making it difficult to compare them. 
We hereby perform a test on the recent \NH\, simulation which has spatial and mass resolutions that are remarkably high for a large-volume simulation, to resolve the situation. 
We generate mock images for the simulated galaxies using SKIRT that calculates complex radiative transfer processes in each galaxy. 
We measure morphological and kinematic indicators using photometric and spectroscopic methods following observer's techniques. 
We also measure the kinematic disk-to-total ratios using the Gaussian mixture model and assume that they represent the true structural composition of galaxies.
We found that spectroscopic indicators such as \vsig\, and $\lambda_{\rm R}$ closely trace the kinematic disk-to-total ratios. 
In contrast, photometric disk-to-total ratios based on the radial profile fitting method often fail to recover the true kinematic structure of galaxies, especially for small galaxies. 
We provide translating equations between various morphological indicators. 
\end{abstract}


\section[]{Introduction}
\label{sec:introduction}
In the era of modern astronomy, various observations have been made with regard to external galaxies. 
From the local to the high-redshift Universe, many observations have suggested that galaxies exhibit various morphological properties. 
Such diversity indicates that galaxies comprise a complex mixture of kinematic components rather than a single structure.
Multiple structures may provide an essential clue to the formation process of galaxies. 
Based on this idea, an independent field of study exists for morphological classification \citep[][]{Hubble1926,SP1940,Holmberg1958,DV1959,vdB1960,Sandage1975}.
The simplest exercise is based on visual inspection.  
However, with the improvement of the surface photometric technique we can now perform radial profile fitting for the structural decomposition of the galaxy often using \sersic\ parameterization \citep[][]{Sersic63}. 

The profile fitting results suggest the presence of ``disk'' and ``bulge'' components in a galaxy, which provides the most widely used morphology index, i.e., the bulge-to-total ratio ($B/T$) or disk-to-total ratio ($D/T$). 
In addition, an extended stellar halo and/or an extra component in the nuclear region is often implied.
Disks are often expressed with the $n = 1$ \sersic\ index, that is, the so called exponential disk.
Classical bulges are generally referred to as \sersic\ with $n = 4$ components, whereas pseudo-bulges show $n < 2$ \citep[][]{Carollo97,Kormendy06}.
Photometric decomposition is usually performed using open-source software tools: e.g., \IRAF, \GALFIT\ \citep[][]{Peng10}, or \PROFIT\ \citep[][]{Robotham17}.
While photometric decomposition is often based on profile fitting, \citet{Zhu18b,Santucci22} attempted to conduct kinematic decomposition using integral field spectroscopic data, e.g., \CALIFA\ and \SAMI\ \citep[][]{GD15,SFSanchez16,vdSande17,Rawlings20}.

In numerical simulations, galaxy properties, including morphology, are sensitive to the cosmology adopted mainly because of the dark matter and thus galaxy assembly history is determined by cosmology. Moreover, understanding astrophysical processes, such as stellar and AGN feedback, is crucial for generating the galaxy morphology distribution realistically \citep[][]{Ubler14}. 
As we possess a ``concordant'' cosmological understanding of the Universe \citep[][]{Komatsu11} and a reasonable consensus for the feedback effect, it is an urgent task to see if  state-of-the-art simulations reproduce the critical properties of galaxies, in particular, morphology. 
Indeed, recent simulations have provided an array of beautiful and seemingly-realistic images of galaxies \citep[][]{Vogelsberger14,Schaye15,Dubois16,Pillepich19,Dubois21}. 

Different techniques are used to determine the morphology of galaxies through simulations.
For instance, a widely-used property is the ratio between the speeds of ordered (rotating) motion and random motion, that is, \vsig\ or kappa parameter (\kppar) \citep[][]{Sales10}.
These parameters may be effective for separating early- from late-type galaxies in a large sample; however, the demarcation cut is still arbitrary.
For example, \citet[][]{Dubois16} found that a cut of \vsig\ $ = 1$ roughly satisfied the observed fractions of early and late-type galaxies.

The circularity parameter is another way to measure the degree of rotational dominance of a galaxy.
\citet[][]{Abadi03} introduced it using the angular momentum and the binding energy of the stellar particles in a galaxy.
As the circularity parameter is measured for all the star particles of the simulated galaxy, we can express the ``kinematic morphology'' based on the circularity distribution. 
Thus, the use of circularity distribution may reduce the degree of uncertainty compared to \vsig\ but is still  subject to the same problem of arbitrariness.
An important concern is how well visual or morphological classifications trace the kinematics of galaxies (\vsig\ or circularity).
\citet[][]{Scannapieco10} have indeed demonstrated that the morphology indicator $D/T$ from photometry and kinematics based on circularity do not agree well with each other for their simulated galaxies.
This is a severe issue for the galaxy community.
Their claim was based on eight simulated galaxies with spatial resolutions of roughly 1~kpc.
We aim to verify this claim using a larger sample of galaxies based on a more up-to-date simulation with much higher resolution.

All these issues originate from one question. Can we identify the pivotal structures of galaxies distinctively and reliably? 
With its high spatial and mass resolution, the hydrodynamical cosmological simulation NewHorizon \citep[][]{Dubois21} is a excellent testbed for answering this question.
To directly compare with observational data, we generate mock images using \SKIRT, a radiative transfer pipeline \citep[][]{CB15,CB20}. 
We measure the weights of the disk and spheroid components following the standard observational technique.  
We compare the photometric and kinematic values of $D/T$ of every galaxy with a stellar mass of over $10^{9.5}\,\rm M_{\odot}$ and attempt to find or quantify the correlation between them. 
We then discuss the degree of (dis)agreement between the two. Finally, we provide a translator between observation and the \NH\ simulation.


\begin{figure*}
\centering
\includegraphics[width=0.8\textwidth]{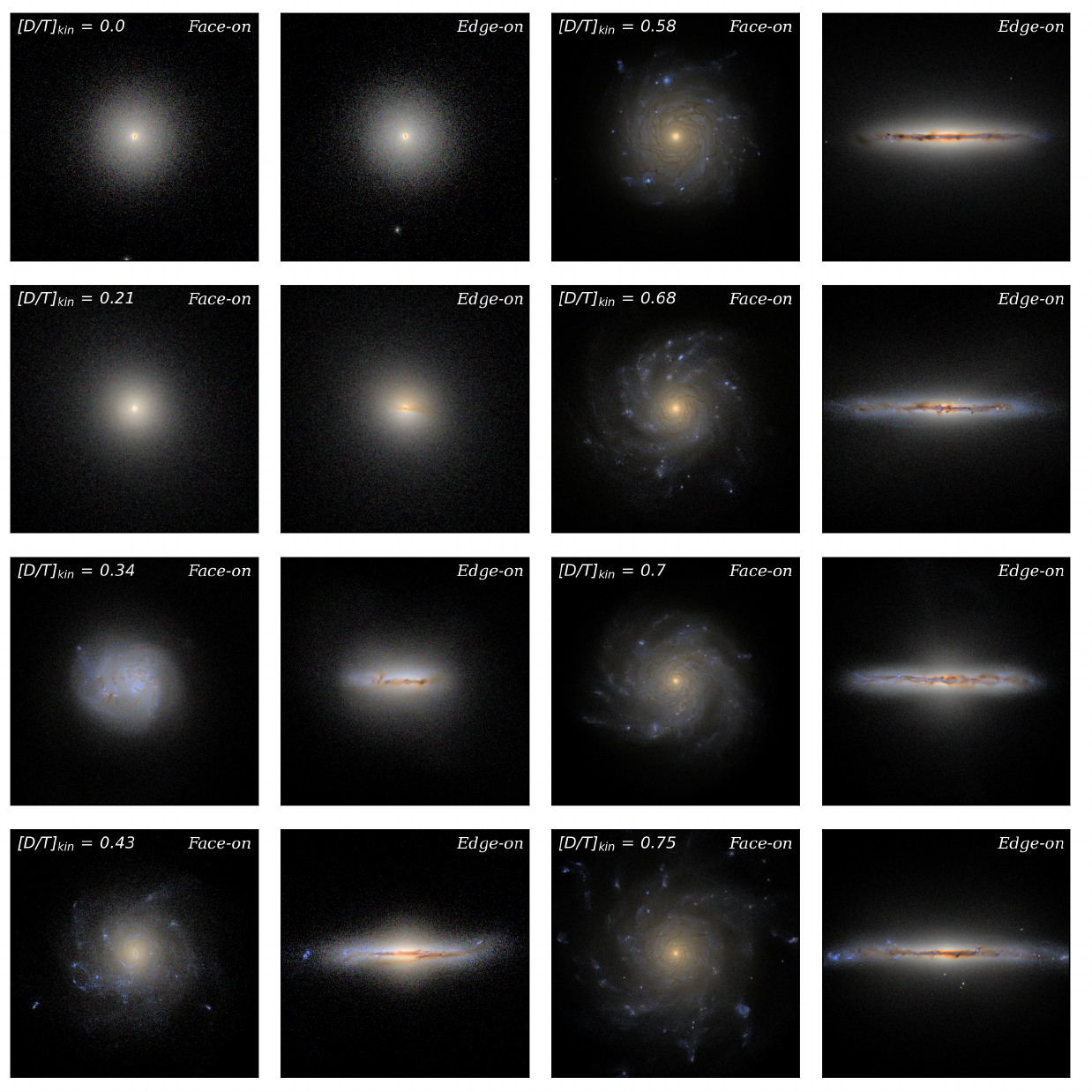}
\caption{
Eight sets of the face-on and edge-on mock images created using the \SKIRT\ pipeline for various values of disk-to-total ratios. 
Red, green, and blue colors are corresponding to SDSS $i$-, $r$-, and $g$-band fluxes, respectively.
The kinematic disk-to-total ratios (see Section~\ref{sec:kinematic decomposition} for definition) of the sample galaxies are given in the face-on images. 
}
\label{fig:Figure1}
\end{figure*}


\section{Methodology}
\label{sec:methodology}

\subsection{The Sample}
We use \NH\ \citep[][]{Dubois21}, a high-resolution cosmological hydrodynamic zoom-in simulation of galaxy formation. 
It covers a spherical ``field'' region in \HAGN\ \citep[][]{Dubois16} with a radius of $10\,{\rm Mpc}$.
Both simulations were performed with \RAMSES\ \citep{Teyssier02}, an adaptive mesh refinement (AMR) code.
The maximum spatial resolution of \NH\ considering the AMR structure is $\Delta x = 34\, \rm{pc}$, and the mass resolutions are $10^{4}$ and $10^{6}\, \rm M_{\rm \odot}$ for stellar and dark matter particles, respectively.
The simulation was executed with the following cosmological parameters, consistent with the WMAP-7 data \citep[][]{Komatsu11}: Hubble constant, $H_{0} = 70.4\,{\rm km\,s}^{-1}\,{\rm Mpc}^{-1}$; total mass density, $\Omega_{\rm m} = 0.272$; total baryon density, $\Omega_{\rm b} = 0.0455$; dark energy density, $\Omega_{\rm \Lambda} = 0.728$; the amplitude of power spectrum, ${\sigma}_{\rm 8} = 0.809$; the spectral index, $n_{\rm s} = 0.967$.
A detailed description of \NH\ can be found in \citet[][]{Dubois21}.

\NH\ contains a substantially smaller number of massive galaxies than the parent simulation, \HAGN, simply because of the volume difference (roughly a factor of 543). 
In order to resolve kinematic structure, we used only the most massive galaxies in the \NH\ simulation. 
To secure a large sample size we used the data from three different snapshots assuming that the morphology and kinematic structure of a galaxy at different snapshots are independently determined.

For galaxy detection, we used the AdaptaHOP algorithm \citep[][]{Aubert04} with the most massive sub-node mode method \citep[][]{Tweed09} for stellar particles. 
For the initial detection of a galaxy, a cut of $N_{\rm ptcl} > 50$ was used, where $N_{\rm ptcl}$ is the number of particles in the detected galaxy candidate.
The center of a galaxy is defined using the density distribution, that is, the position of the stellar particles on the highest density peak. 
We sample the galaxies with the total stellar mass greater than $10^{9.5}\,\rm M_{\rm \odot}$, which corresponds to $N_{\rm ptcl} \gtrsim 3.6 \times 10^{5}$ (75 galaxies at $z=0.7$, 92 galaxies at $z=0.3$ and 107 galaxies at $z=0.17$).
Also, we exclude the irregular or merging galaxy samples (38 galaxies) based on the visual morphology classification. 


\subsection{Mock Imaging}
\label{sec:mock imaging}

For visual inspection and photometric classification following the observer's approach, we first generate the mock images of the \NH\ galaxies using \SKIRT. 
Assuming that a certain number of photon packets are radiated from each light source, \SKIRT\ traces each ray and calculates the attenuation from the gas cells along its path.
We used the ``BC03'' simple stellar population models \citep[][]{BC03} to calculate the light from the sources.
We assume that only the gas cells with the temperature under $10,000\, \rm K$  can have the dust content in it.
We use the dust population model of \citet[][]{Zubko04}, allowing each gas cell to include silicate, graphite, and PAH populations each of which has 15 size varieties.
We estimate the dust abundance of each cell as $M_{\rm dust} = M_{\rm cell} \times Z \times f_{\rm dust}$, where $Z$ denotes the metallicity and $f_{\rm dust}$ denotes the dust-to-metal ratio. We set $f_{\rm dust}$ to 0.3 as a fixed value for every gas cell.
The quality of the image basically depends on the number of photon packets per wavelength.
We assume the number of photon packets to be $8 \times 10^{7} \times (M_{\rm gal}\,/\,10^{10}\,\rm M_{\rm \odot})$ photon packets depending on the galaxy mass, where $M_{\rm gal}$ denotes the total stellar mass of a galaxy.
Other studies have adopted adaptive photon packet number scaling with pixel size \citep[][]{Rodriguez-Gomez19}.
In this study, the pixel scale of the resultant image is \dx\,(i.e., $34\, \rm pc\, \rm pix^{-1}$).
\SKIRT\ can provide the calculation of a second (cascade) radiation, i.e., the thermal radiation from each gas (dust) cell, and characteristic emission lines from star-forming regions; however, they were not considered in our analysis because we focus on optical bands in this study. 

We include seeing effects by adding Gaussian dispersion with a standard deviation of 3 pixel lengths but not the background noise.
The outermost part of the galaxy's surface brightness is obviously more affected by the noise. 
However, with logarithmic radial binning and a strict radius cut for radial profile we can achieve reasonably-robust measurements against the uncertainty in the background noise.
Figure~\ref{fig:Figure1} shows the mock images of eight sample galaxies for various values of disk-to-total ratios.
The color scheme is the same as that of \cite{Lupton04}.

\subsection{Decomposition}
\label{sec:decomposition}


\subsubsection{Kinematic Decomposition}
\label{sec:kinematic decomposition}

\begin{figure}
\centering
\includegraphics[width=0.45\textwidth]{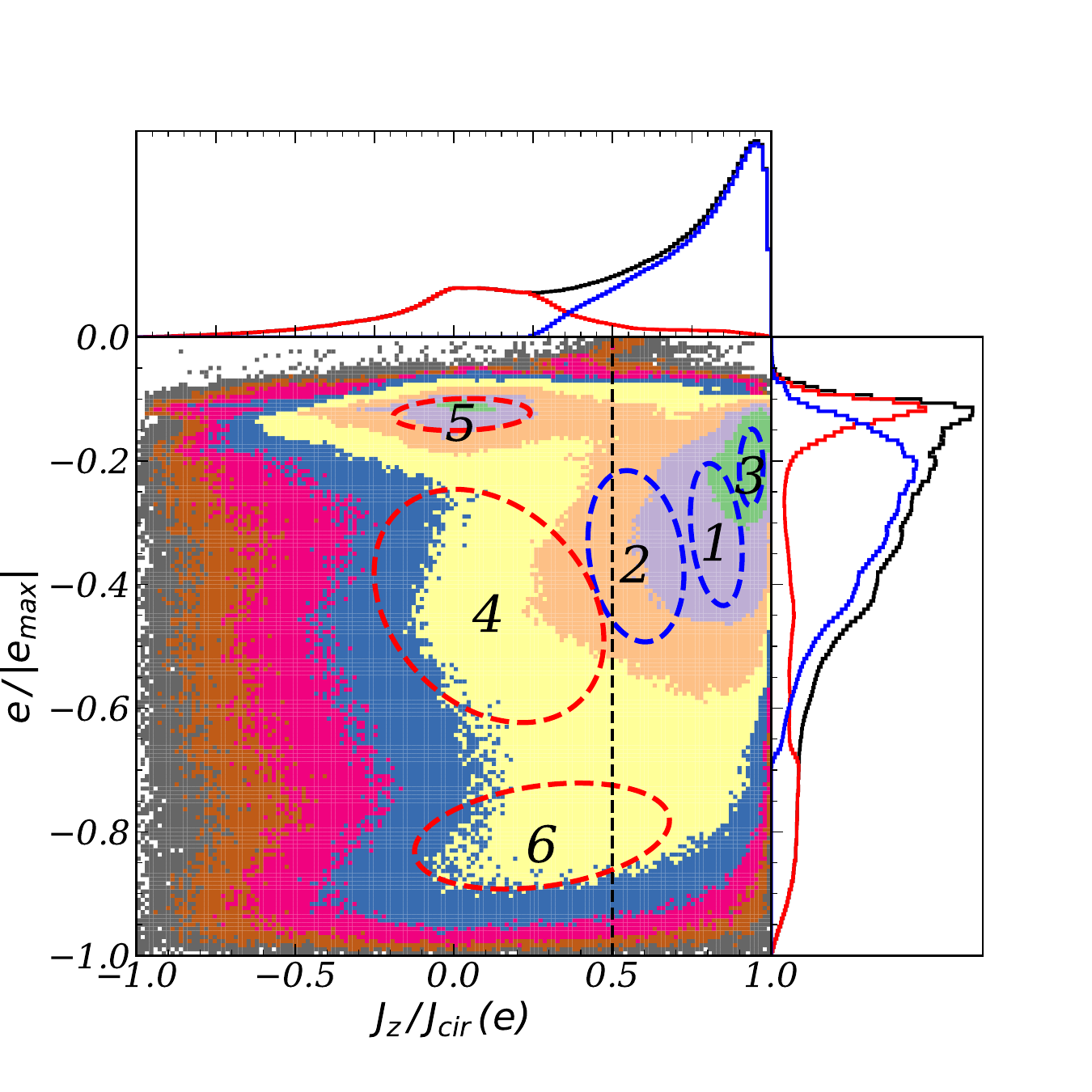}
\caption{
The exemplary phase-space distribution of the disk galaxy in Figure~\ref{fig:Figure1} (second from the top on the right with \dttk\ = 0.68). 
While three parameters are used for component detection as described in the text, we hereby show only the energy and circularity parameters.
The GMM clustering result with N$_{\rm comp}$\,=\,6 is also marked in the figure. Y-axis shows the specific binding energy normalized with the most bound particle's specific binding energy, and X-axis shows circularity parameter. Ellipses show 1-${\sigma}$ of multivariate Gaussian distribution, and the numbers are ordered with their mass weight. 
Red and blue colors correspond to the spheroidal and disk components, respectively.
The panels at the top and the right show their combined projected distributions of the circularity parameter or the specific binding energy.
}
\label{fig:Figure2}
\end{figure}

\begin{figure*}
\centering
\includegraphics[width=0.85\textwidth]{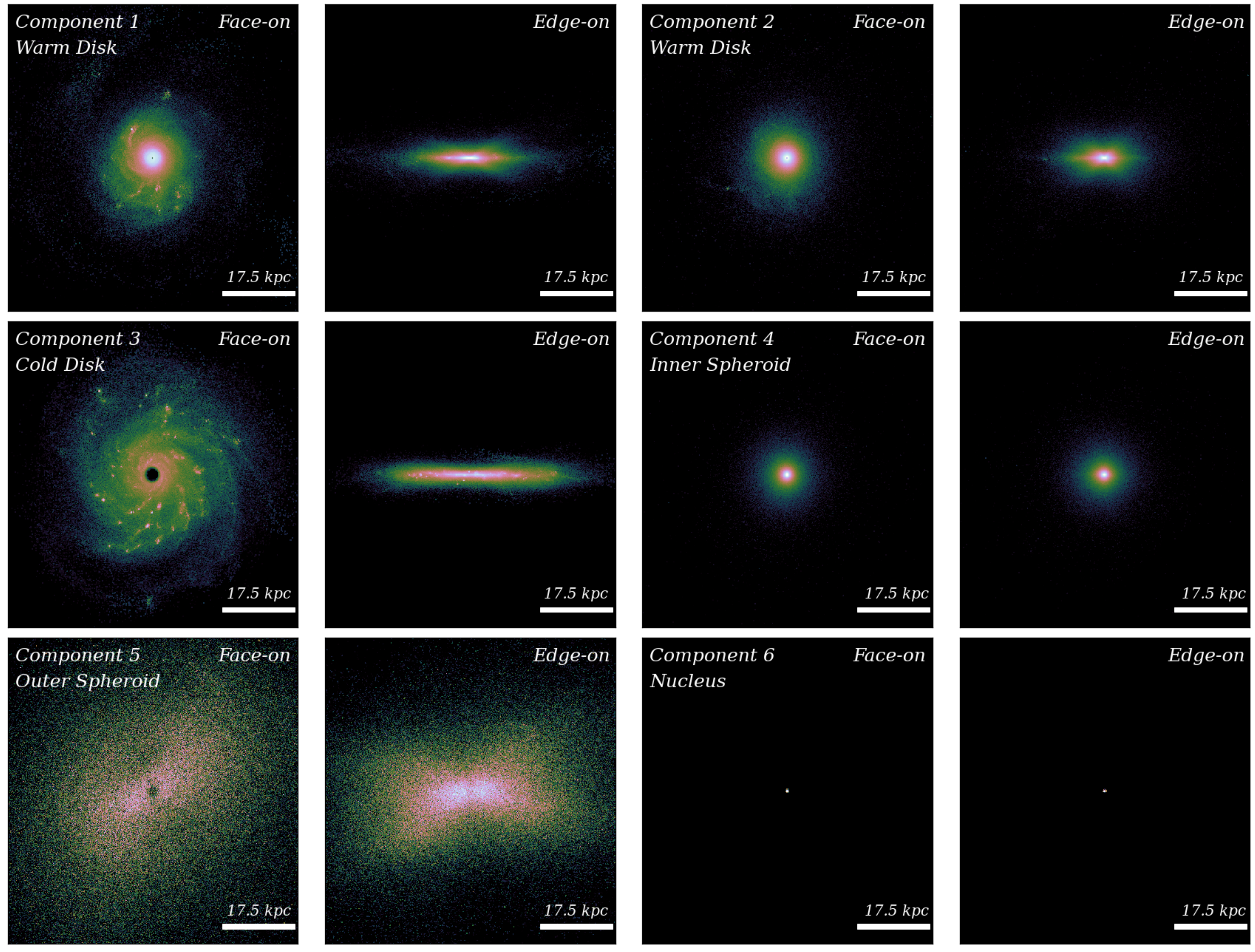}
\caption{
The face-on and the edge-on $r$-band flux density map of each GMM component shown in Figure~\ref{fig:Figure2}. 
}
\label{fig:Figure3}
\end{figure*}

The kinematic decomposition of the \NH\ galaxies was performed based on three key parameters. 
The first is circularity. 
The circularity (${\epsilon}$) of each stellar particle can be defined in two different ways, using the radial distance or the binding energy of a particle.
We adopt the latter. Circularity is defined as
\begin{equation}
\epsilon = J_{\rm z}\,/\,J_{\rm cir}({e}),
\end{equation}
where $J_{\rm z}$ denotes the angular momentum of each star particle along the bulk rotation axis of a galaxy, and $J_{\rm cir}({e})$ denotes the maximum angular momentum that a stellar particle can have with a specific binding energy ($e$).
We assumed a spherically symmetric potential for calculating the binding energy to perform a fair comparison with the circularity parameters derived from spectroscopy \citep[e.g.,][]{Zhu18a,Zhu18b}.
The circularity parameter (${\epsilon}$) is widely used to decompose the kinematic structures of simulated galaxies \citep[see e.g.,][]{Park19,Park20}.
Furthermore, we use two additional parameters for kinematic decomposition as considered in recent studies \citep[e.g.,][]{Obreja18,Du19,Du20}. 
The first is the remaining angular momentum, i.e., \ppar, where $\vec{J_{\rm p}} = \vec{J} - \vec{J_{\rm z}}$. 
The second parameter (namely, ``energy parameter'') is the specific binding energy of a particle normalized by the value of the most bound particle, i.e., \epar. 

Considering these parameters (\cpar, \ppar, and \epar), we can decompose the kinematic structures of a galaxy based on their three-dimensional phase-space distributions.
The specific locations of structures in the phase-space vary especially along the energy parameter axis, depending on galaxies and the presence of a centrally-concentrated component (e.g., bulge).
However, the distribution is so smooth that it is challenging to group (``cluster'') stellar particles into various kinematic components.

We use a Gaussian mixture model (GMM), an un-supervised machine learning clustering technique, to overcome this difficulty.
Recent studies have used GMM to decompose galaxies into detailed structural components \citep[e.g.,][]{Du19,Du20}.
For instance, \citet[][]{Du19} suggests that for the TNG100 simulation \citep[][]{Nelson18}, the mean positions of clusters identified with GMM are located in four different regions in the energy versus circularity space: two for the spheroidal component (${\epsilon}<0.5$, bulge \& halo) and the other two for the (thin and thick) disk components (${\epsilon}>0.5$) (refer to Fig. 3 in \citeauthor{Du20} \citeyear{Du20}).
We apply GMM to \NH\ galaxies.
The number of components is a free parameter in GMM, and we have tried various numbers up to 15.
We set it to be 6 ($N_{\rm comp}\,=\,6$) in this study so that the comparison with observations becomes simple and intuitive.
To be more specific, we wanted to detect the structural components that observers often detect and discuss: e.g., thin and thick disks for rotating components, and bulge, inner and outer halos for dispersion components. 
In order to detect these five components using GMM, the minimum value of $N_{\rm comp}$ was found to be 6 in most cases.

Figure~\ref{fig:Figure2} shows the GMM ``components'' in the energy-circularity space for a sample galaxy. 
Each component is assigned an ID in the order of mass weight, where 1 corresponds to the highest weight.
The detailed spatial distributions of the components are presented in Figure~\ref{fig:Figure3}. 
We classify the six components into five structural components that observers often refer to:
the warm disk (components 1 and 2), cold disk (component 3), inner spheroid (component 4), outer spheroid (component 5), and nucleus (component 6), although the direct comparison between the GMM-detected structures and observational structures may not be straightforward.

We note that 38\% of our galaxies possess a kinematically distinct component near the galactic center.
Their size (the mean distance of all the star particles of the component) is roughly 12\% of the half mass radius (\rft) of the galaxy, or 0.2~kpc, and their mass fraction is about 11\%.
Since they appear to be more centrally concentrated than typical bulges, we call them a ``nucleus'' in this study.
If it is real, conventional bulges may be a combination of the nucleus and (part of) the inner spheroid.
The detailed distribution of star particles in the phase space varies substantially from galaxy to galaxy. We present the phase-space
diagram of other galaxy in Appendix for reference. 
The existence of such a nucleus in galaxies as a kinematically-distinct structure is an interesting issue. It will be a subject of future investigations.

\begin{figure}
\centering
\includegraphics[width=0.45\textwidth]{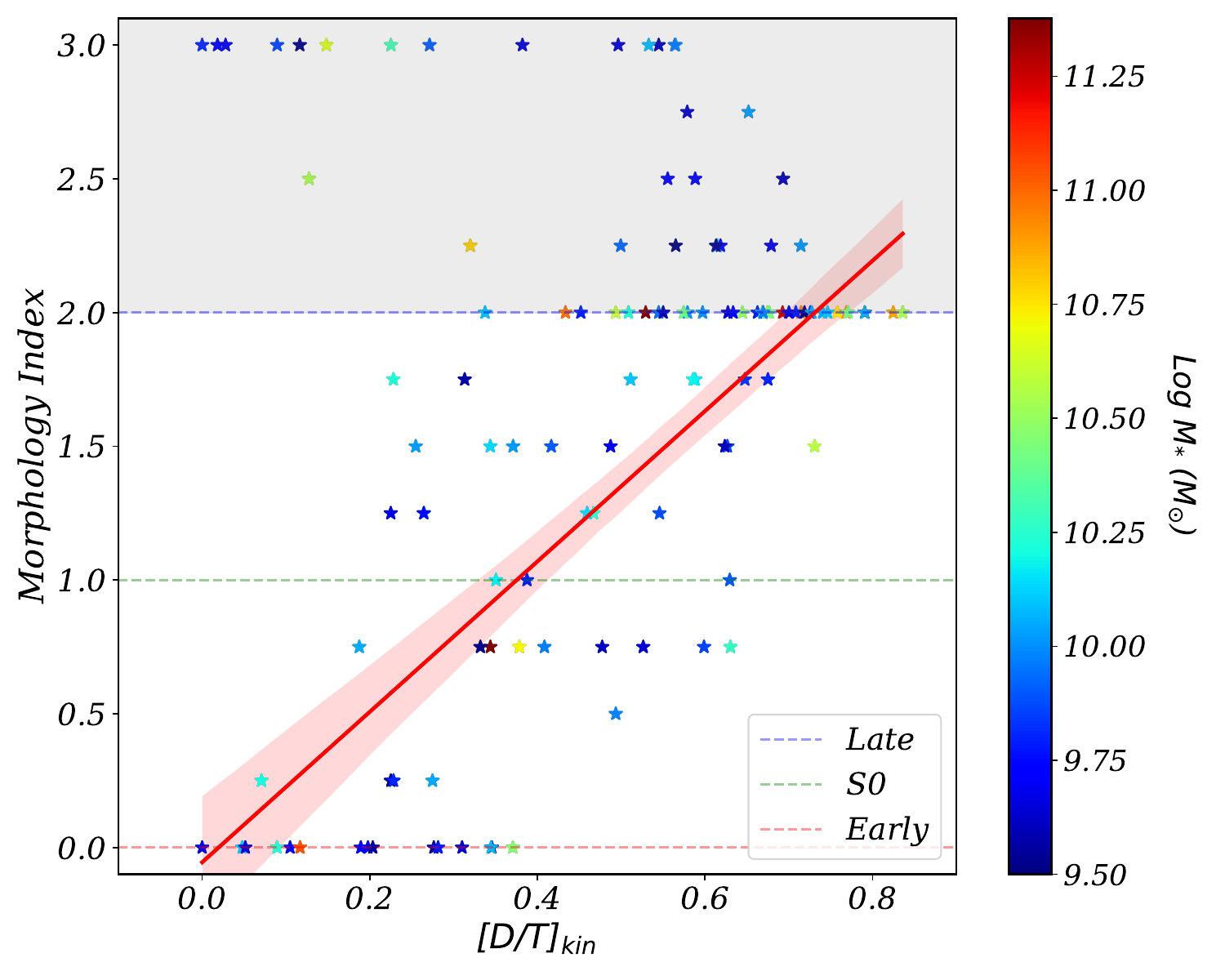}
\caption{
Visual morphology versus kinematic disk-to-total ratio for the sample galaxies at $z=0.17$.
Visual morphology classification was performed by four members of the authors, and each index value is a mean value of the four.
Individual members classified galaxies into early-type (index 0), lenticular (1), late-type (2), and unclear (3) galaxies.
Except for the unclear galaxies (morphology index $>2$), visual morphology and \dttk\, shows a good correlation (Pearson correlation coefficient r $\sim$ 0.80).
The red line with shade shows the linear fit and 1-$\sigma$ error.
The color key on the side shows the stellar mass information for the galaxies.
}
\label{fig:Figure4}
\end{figure}

The kinematic disk-to-total ratio, \dttk, is defined as the mass ratio between the combined mass of the disk components 
and the total stellar mass:
\begin{equation}
[D/T]_{\rm kin} = M_{\rm disk}\,/\,M_{\rm total}.
\end{equation}
where $M_{\rm disk}$ is a total mass of GMM components with mean circularity $\bar{\epsilon}\, >\, 0.5$.
We measure both $M_{\rm disk}$ and $M_{\rm total}$ inside \rnt, where \rnt\ denotes the radius within which 90\% of the total stellar mass resides.
We assume \dttk\, as ``ground truth'', a representative structural property of a galaxy in this study. 
We compare it with the visual morphology based on the face-on and the edge-on projection determined by four members of the authors in Figure~\ref{fig:Figure4}. 
We used an arbitrary digital scheme: 0 for early-type, 1 for lenticular, 2 for late-type, and 3 for unclear type. 
The Pearson correlation coefficient ($r$) was measured in the range of 0 to 2, as 3 for the unclear type is an irrelevant value.
The correlation is reasonably good with $r$ of 0.80, thereby confirming that the kinematic disk-to-total ratio agrees with the visual morphology.


\subsubsection{Photometric Decomposition}
\label{sec:photometric decomposition}

\begin{figure*}
\centering
\includegraphics[width=0.7\textwidth]{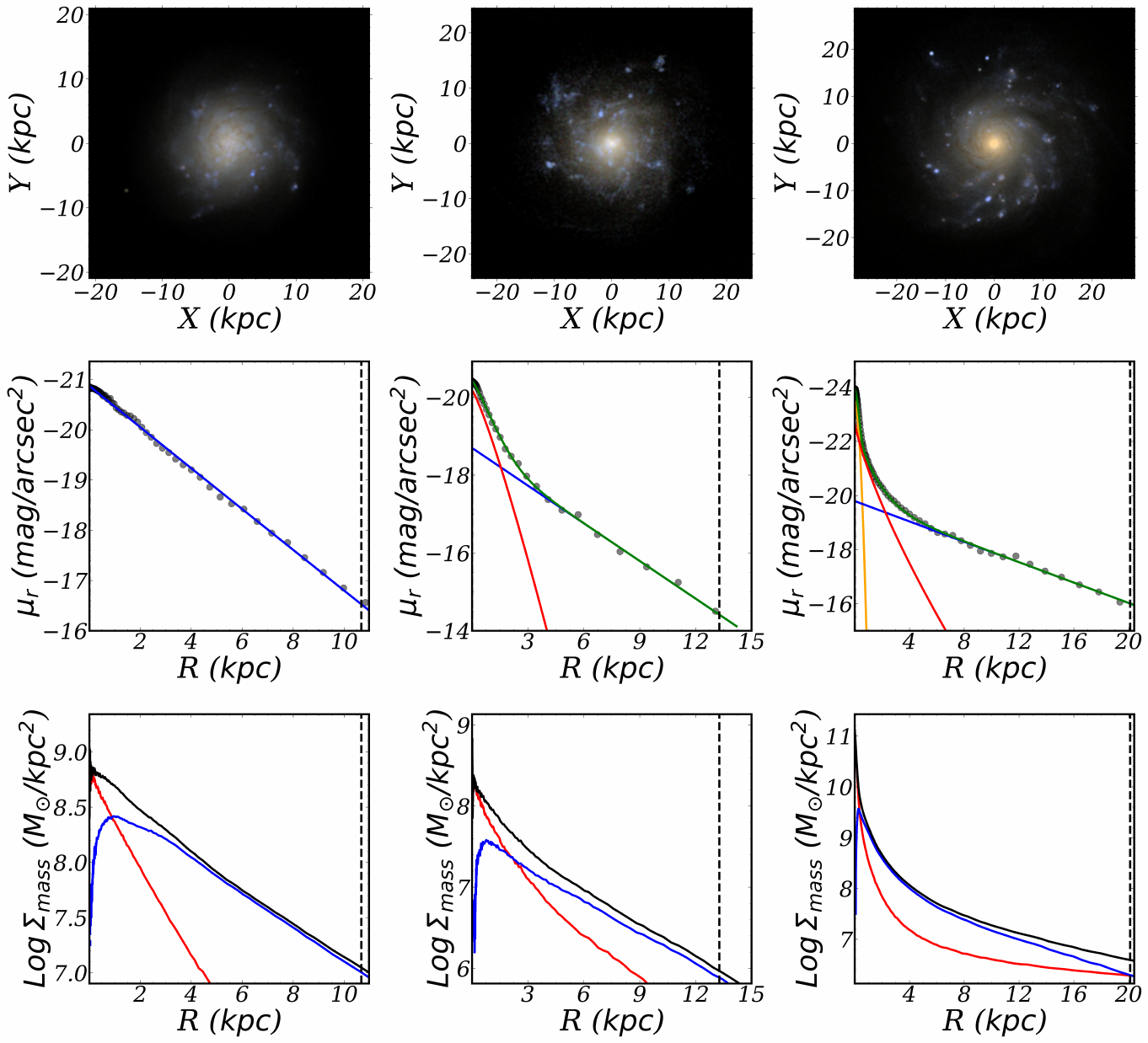}
\caption{
Examples of photometric decomposition result. 
The galaxy in the right column is the one as shown in Figure~\ref{fig:Figure2}.
Each column shows the optimal fitting result based on BIC, best-fitted by one, two, and three component respectively. 
The first row shows the face-on mock image of a galaxy created using sdss g, r, i flux. 
The second row shows the surface brightness profile fitting result. 
The disk (blue), inner spheroid (red), nucleus (orange), and total galaxy (green)'s surface brightness profiles are shown. 
The last row shows the surface mass profile.
The disk (blue) and spheroid (red) are divided using the kinematic decomposition as was done for the circularity distribution histogram in Figure~\ref{fig:Figure2} (top panel), and the total (black) surface density profile is the sum of the two.
The black dashed vertical lines are the \rnt\ of the galaxies.
}
\label{fig:Figure5}
\end{figure*}

\begin{figure*}
\centering
\includegraphics[width=0.9\textwidth]{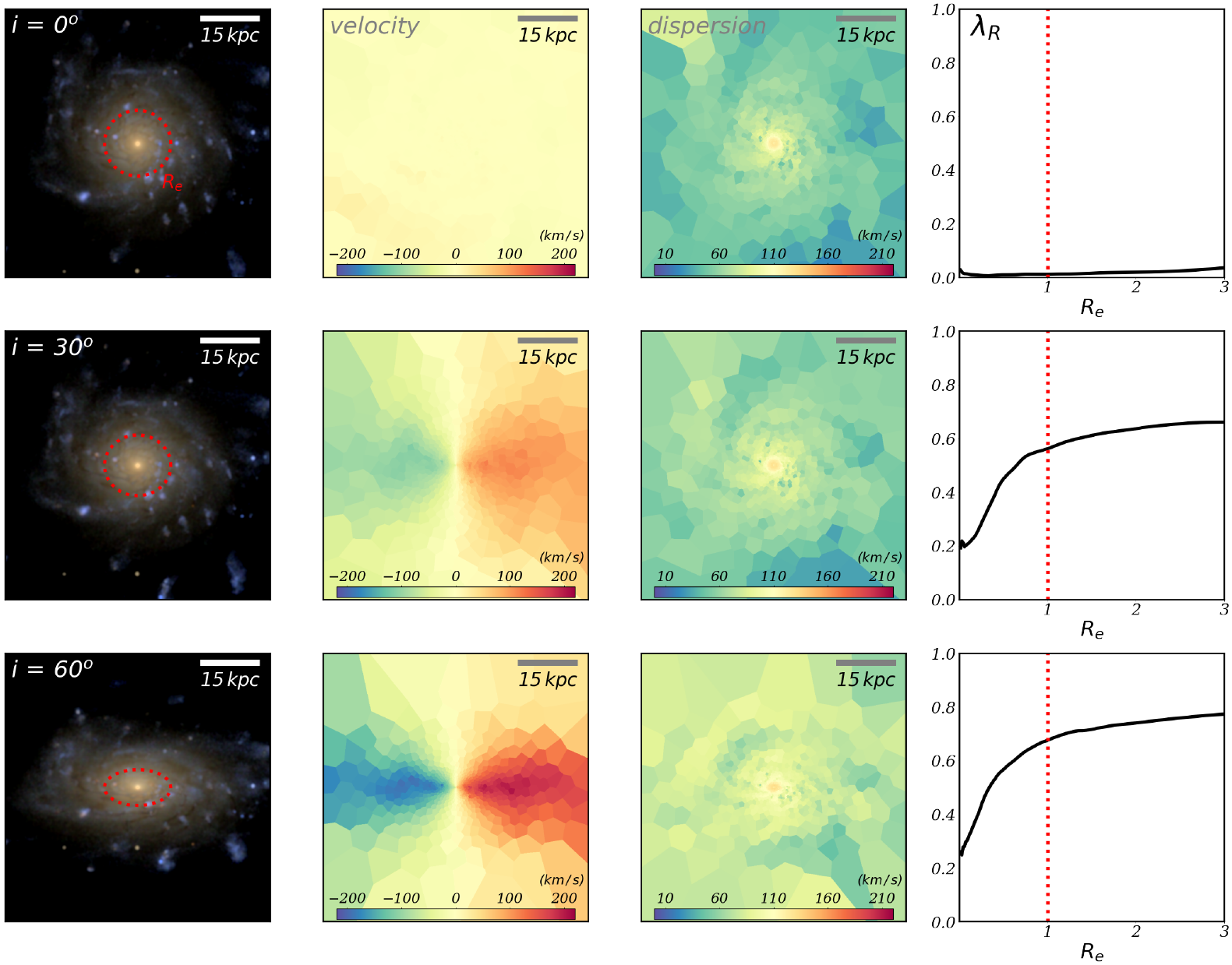}
\caption{
Spin parameter estimation for a galaxy with different inclinations: 0, 30, 60 degree for each row. 
The first column shows the mock images of the galaxy based on the SDSS $g$-,$r$-,$i$-band magnitudes. 
The second and the third columns are a velocity and dispersion maps. 
The last column is the spin parameter measured as a function of radial distance. 
The red dashed line indicates the effective radius. 
}
\label{fig:Figure6}
\end{figure*}

We use the mock images described in Section~\ref{sec:mock imaging} for photometric decomposition.
We assume that each photometric component follows \sersic\ profile \citep[][]{Sersic63}: 
\begin{equation}
I(R) = I_{\rm e} \, \exp(-b_{\rm n}\,[(R\,/\,R_{\rm e})^{1/n}\,-\,1]),
\end{equation}
where $R_{\rm e}$ denotes the effective (half-light) radius, $I_{\rm e}$ the luminosity at $R_{\rm e}$,  $n$ the \sersic\ index, and $b_{\rm n}$ a function that depends on the \sersic\ index. 
The \sersic\ profile is a well-known model that can express both the disk and bulge components of galaxies with different \sersic\ index values ($n$), conventionally 1 for the former and 4 for the latter.
According to a recent study, using a single exponential profile may over-predict the disk component in the inner part of galaxies \citep[][]{Papaderos22}.
However, due to the overall wellness of the fit and the difficulty of applying a more complicated model to observed profiles that are often of limited quality, the majority of observations and surveys still use \sersic\ profile fitting.

For the photometric decomposition of the \NH\ galaxies, we assume that galaxies can have up to four components: ``nucleus'', ``inner spheroid'', ``disk'', and ``outer spheroid''.
We do not consider tidal features in the fitting procedure.
We fit the profile with a one-dimensional surface brightness radial profile. 
We use circular apertures for profile fitting mainly because we only use the face-on images of galaxies.
To simplify the analysis and interpretation, we set the \sersic\ index of the disk component to 1. 
We treat the \sersic\ index as a free parameter for the other components.
For the radial extension, we use the data inside \rnt.
For our sample, \rnt\ is $2.70^{+1.03}_{-0.40}$ times the \rft\ median.
We performed profile fitting to rest-frame $r$-band images.
We assume that galaxies are at a distance of 1\,Mpc regardless of the redshift; and thus considering the high resolution of \NH, our images are of good quality in terms of surface brightness per pixel.

While all the four components are used for radial fits, we select the best model based on the Bayesian information criterion (BIC) as follows:
\begin{equation}
\mathrm{BIC} = -2\,\mathrm{Log}\,\mathcal{L}\,+\,k\,\mathrm{Log}\,n_{\rm dat},
\end{equation}
where $\mathcal L$ denotes the likelihood of fit, $k$ denotes the number of parameters, and $n_{\rm dat}$ denotes the number of data points.
In practice, none of our galaxies needed all four components for a BIC-based best fit. 
Approximately 34\% (80 out of 236) of galaxies were best fitted as single-component disk galaxies, i.e., ``pure disks''.
29\% of galaxies were best fitted by a combination of two components (disk and spheroid).
The rest (37\%) required an additional ``nucleus'' component.
Approximately 54\% of the three-component fitted galaxies were found to possess a nuclear component based on the GMM kinematic decomposition in Section~\ref{sec:kinematic decomposition}.
Figure~\ref{fig:Figure5} shows galaxies best-fitted by one, two, or three components. 
We provide in the bottom row the radial {\em mass} profile of the galaxies derived from the GMM analysis for reference.

We defined the photometric disk-to-total ratio as the luminosity fraction of the disk component as follows:
\begin{equation}
[D/T]_{\rm phot} = L_{\rm disk}\,/\,L_{\rm total}
\end{equation}
and we compare this parameter with the result from the kinematic decomposition described in Section~\ref{section:result}. 


\subsection{Spectroscopic Parameters}
\label{sec:spin parameter : method}

We measure spectroscopic parameters to assess the degree of correlation with intrinsic kinematics.
For this, we use two parameters, $\lambda_{R}$ and $V/{\sigma}$.
We used the definition of the spin parameter given in \citet[][]{Emsellem07}.
\begin{equation}
\lambda_{\rm R} = \frac{\Sigma_{\rm i}F_{\rm i}R_{\rm i}|V_{\rm i}|}{\Sigma_{\rm i}F_{\rm i}R_{\rm i}\sqrt{V_{\rm i}^{2}+\sigma_{\rm i}^{2}}}.
\end{equation}
where $F_{\rm i}$, $R_{\rm i}$, $V_{\rm i}$, and $\sigma_{\rm i}$ are the attenuated r-band flux, distance from the center, line-of-sight (los) velocity, and los velocity dispersion of i-th spaxel.

We use the same axis ratio of the ellipse measured at \reff\ in the SDSS $r$-band flux for all the radial bins.
For each inclination, we generate velocity and dispersion maps using the Voronoi tessellation algorithm based on \citet[][]{CC03} to ensure that the $S/N$ ratios of the bins are virtually the same.
At least five stellar particles are present in each bin.
Figure~\ref{fig:Figure6} shows an example of the measurement. 
The first column exhibits RGB color images using the SDSS $i$, $g$, and $r$-band fluxes. 
The red dashed ellipse indicates the ellipse fit at \reff.
The second and third columns show the velocity and velocity dispersion maps, respectively.
The last column shows the spin parameter measurements inside 3\,$R_{\rm e}$. 
The rows demonstrate the results for different inclinations.

The rotation-to-dispersion ratio, \vsig, was measured also considering flux weights and within \reff\ using the following definition: 
\begin{equation}
(V/\sigma)^{\rm 2}\ =\ \frac{\Sigma_{\rm i}F_{\rm i}V_{\rm i}^{\rm 2}}{\Sigma_{\rm i}F_{\rm i}\sigma_{\rm i}^{\rm 2}},
\end{equation}
where $F_{\rm i}$, $V_{\rm i}$, and $\sigma_{\rm i}$ are same as above.
We choose different inclinations to measure two spectroscopic parameters, $\theta$ = $30^{\circ}$, $60^{\circ}$, and $90^{\circ}$. 

\begin{figure}
\centering
\includegraphics[width=0.4\textwidth]{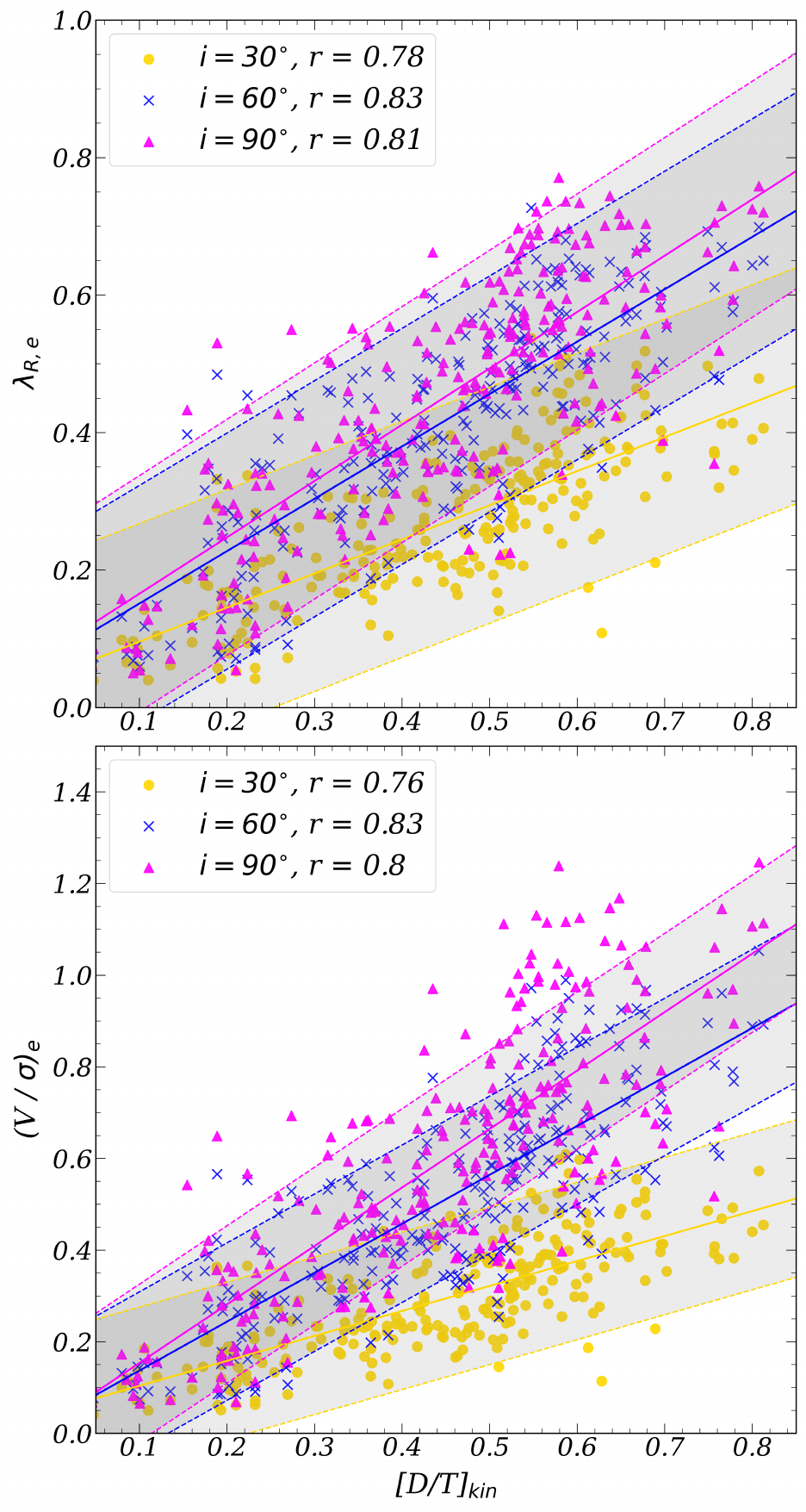}
\caption{
The relation between the spectroscopic parameters and the kinematic disk-to-total ratio, $[D/T]_{\rm kin}$. 
The upper panel shows the relation for \spinp, and the lower panel shows for $V/{\sigma}$.
The lines with shades show the linear fit and 1-$\sigma$ error to the data of different inclinations.
The general properties of correlations
are given in Table~\ref{tab:Table1}.
}
\label{fig:Figure7}
\end{figure}

\begin{figure}
\centering
\includegraphics[width=0.4\textwidth]{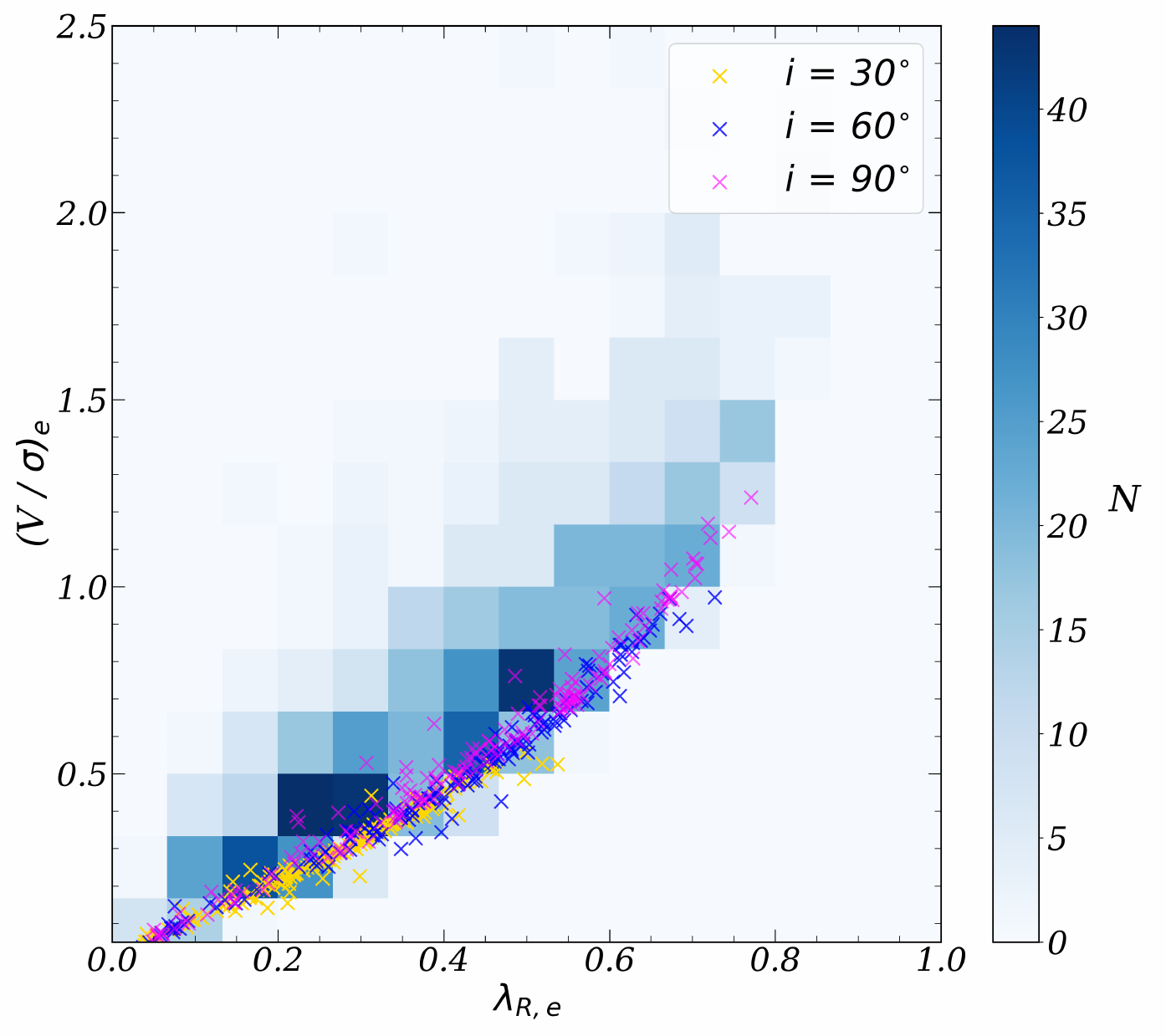}
\caption{
The spectroscopic spin parameters of the \SAMI\ observational data (Hess diagram) and the \NH\ simulation data (crosses). 
}
\label{fig:Figure8}
\end{figure}


\section{Results}
\label{section:result}

In this section, we assess the validity of various morphological and kinematic indicators by comparing them with the kinematic disk-to-total ratio.


\subsection{Spectroscopic parameters}
\label{subsection:spectroscopic_result}

Figure~\ref{fig:Figure7} shows the correlation of the spin parameter $\lambda_{\rm R}$\ (upper panel) and \vsig\ (lower panel) against the kinematic $D/T$.
The three different symbols represent three different inclinations: $30^{\circ}$, $60^{\circ}$, and $90^{\circ}$. 
The linear regression for each inclination is shown with 1-${\sigma}$ standard deviation. 
We found a reasonably good agreement between the spectroscopic parameters and \dttk. 
The general properties of linear regression are given in Table~\ref{tab:Table1}.
This confirms that the spectroscopic spin parameters trace the kinematic structure of galaxies well.

For validation, we compare the \SAMI-observed data \citep[][]{vdSande17b} with the \NH\ simulation galaxies in the plane defined by the two spectroscopic spin parameters, \vsig\ and \spinp, in Figure~\ref{fig:Figure8}. They indeed correlate well with each other. 
The simulated galaxies occupy a region similar to that captured based on the observed data in this plane.
Therefore, it is safe to use spectroscopic spin parameters to extract the kinematic structure or morphology of galaxies.

\begin{table}
  \centering
  \caption{Correlation between parameters and $[D/T]_{\rm kin}$ ($Y$ = $a$\,${[D/T]}_{\rm kin}$\,+\,$b$)}
  \begin{tabular}{ccccccc}
  \hline \hline 
$Y$ & ${\theta}$\,(or\,M$_{\rm gal}$) & $a$ & $b$ & $r-value$  \\ [5pt]
  \hline
 ${\lambda}_{\rm R,e}$ & 30$^{\circ}$ & 0.50 & 0.05 & 0.78 \\ [3.5pt]
 ${\lambda}_{\rm R,e}$ & 60$^{\circ}$ & 0.76 & 0.07 & 0.83 \\ [3.5pt]
 ${\lambda}_{\rm R,e}$ & 90$^{\circ}$ & 0.82 & 0.08 & 0.81 \\ [3.5pt]
 \hline 
 $(V/{\sigma})_{\rm e}$ & 30$^{\circ}$ & 0.54 & 0.05 & 0.76 \\ [3.5pt]
 $(V/{\sigma})_{\rm e}$ & 60$^{\circ}$ & 1.07 & 0.03 & 0.83 \\ [3.5pt]
 $(V/{\sigma})_{\rm e}$ & 90$^{\circ}$ & 1.28 & 0.03 & 0.80 \\ [3.5pt]
 \hline 
 ${[D/T]}_{\rm phot}$ & $> 10^{9.5}$ & 0.51 & 0.55 & 0.35 \\ [3.5pt]
 ${[D/T]}_{\rm phot}$ & $\&\ w/o\ \rm PD^{\rm a}$ & 0.54 & 0.42 & 0.42 \\ [3.5pt]
 ${[D/T]}_{\rm phot}$ & $> 10^{10.3}$ & 0.67 & 0.41 & 0.65 \\ [3.5pt]
 ${[D/T]}_{\rm phot}$ & $\&\ w/o\ \rm PD$ & 0.74 & 0.36 & 0.73 \\ [3.5pt]
 \hline \hline 

 \end{tabular}
 \label{tab:Table1}
 
 \raggedright
 \tablenotetext{a}{PD : pure disk galaxy}
 
\raggedright
\end{table}

\begin{figure}
\centering
\includegraphics[width=0.4\textwidth]{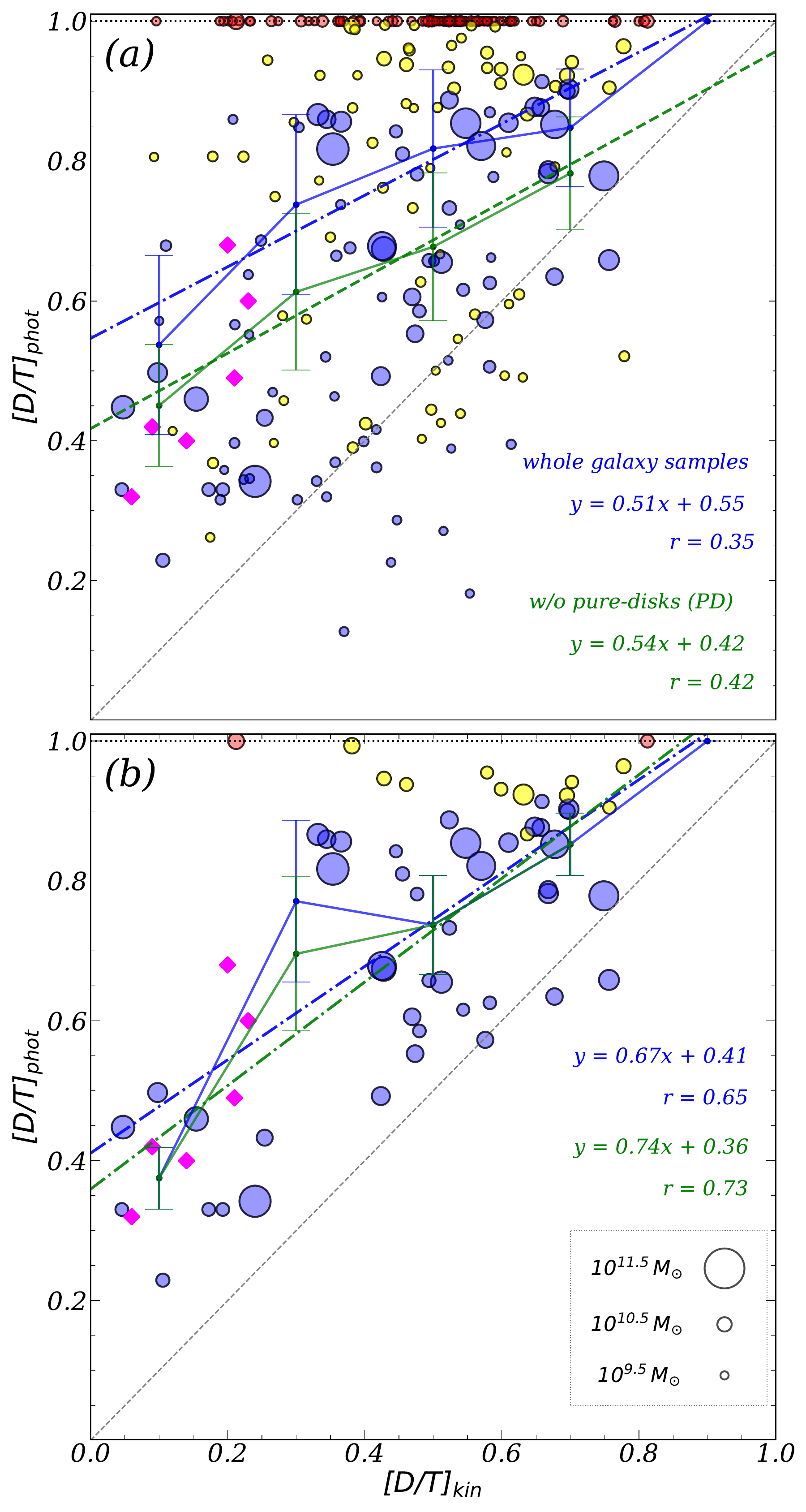}
\caption{The comparison between \dttp\ and \dttk.
The circles denote the galaxies with stellar mass greater than 10$^{9.5}$ M$_{\rm \odot}$ (upper panel) and 10$^{10.3}$ M$_{\rm \odot}$ (lower panel). 
The sizes of the circles are scaled with their mass, between 10$^{9.5}$ and 10$^{11.5}$ M$_{\rm \odot}$.
The blue dashed line is a linear regression for the whole sample, while the green dashed line is for the galaxies with 0.01 $<$ $[D/T]_{\rm phot}$ $<$ 0.99, that is excluding ``pure disks''. 
The magenta diamonds are the data from \citet[][]{Scannapieco10}.
The correlation and the Pearson coefficient are also shown in each panel. 
The galaxies photometrically best-fitted by one, two, or three components are color-coded as red, yellow, and blue, respectively.
}
\label{fig:Figure9}
\end{figure}


\subsection{Photometric parameter}
\label{subsection:photometric result}

Figure~\ref{fig:Figure9} shows our galaxy sample on the \dttp\ versus \dttk\ plane. 
The symbol size was scaled considering the galaxy's stellar mass.
Figure~\ref{fig:Figure9}-(a) shows the entire sample of galaxies above $10^{9.5}\,$\msol.
The 1-${\sigma}$ errors were estimated in five equally-spaced \dttk\ bins.
We also present the 8 simulated galaxies from \citet[][]{Scannapieco10} for comparison.
Our data based on the \NH\ galaxies appear to be compatible to the results of \citet[][]{Scannapieco10}.

As shown in the figure, there is a large scatter, thus leading to a poor (or, at best ``modest'') correlation.
Assuming that \dttk\, traces the true structure of galaxies, \dttp\ does not properly recover it: the correlation coefficient is 0.35 (given inside Panel-a).
Besides, \dttp\ systematically overestimates the disk-to-total ratio.
These facts make it difficult to directly compare the simulated with the observed galaxies.

Furthermore, there are a number of galaxies with an extremely high value of \dttp\ (essentially 1), although their kinematic disk-to-total ratios are substantially lower.
The profile fitting technique classifies them as ``pure disks" whereas GMM clearly detects a substantial dispersion component.
This happens more often for lower mass galaxies: note the small symbol sizes for galaxies with \dttp\,$\approx 1$.
This demonstrates that the profile fitting technique tends to perform inaccurate decomposition for small galaxies.
If we assume that galaxies classified as pure disks are incorrectly done so, we may want to exclude them and re-estimate the correlation between \dttk\, and \dttp.
In this case, we derived a slightly improved correlation coefficient (0.42). 
The translation relations with and without pure disk galaxies are shown in the figure. 
The presence or abundance of massive pure disks is an important issue in terms of cosmology \citep[][]{Kormendy10,Peebles15}, and it appears that photometric fits tend to incorrectly classify galaxies as pure disks. 
This issue deserves further investigations. 

The mis-classified galaxies being typically small and low-mass, we tried various values of mass cut in the range of $M_{\rm cut} = 10^{9.5 - 10.5}$ \msol.
The correlation coefficient monotonically increases with increasing mass cut until $M_{\rm cut} = 10^{10.3}$ \msol\, and dramatically drops beyond that mainly due to the decrease in the sample size.
Therefore, we have decided that the most representative translation between \dttp\, and \dttk\, can be achieved with $M_{\rm cut} = 10^{10.3}$ \msol. 
We show the result in Panel-b of Figure~\ref{fig:Figure9} and give the correlation between the two parameters for a mass cut of $M_{\rm cut} = 10^{10.3}$ \msol\ as follows:
\begin{align}
[D/T]_{\rm phot} &= 0.67\,[D/T]_{\rm kin} + 0.41\ \rm (all) \\
                 &= 0.74\,[D/T]_{\rm kin} + 0.36\ \rm (exc.\ PD)
\end{align}
with a correlation coefficient of 0.65 and 0.73 when pure disks are included or excluded, respectively.
It remains to be tested how galaxies simulated with different physics and numerical approaches distribute along this relation.


\subsection{Discussion}
\label{subsection:discussion}

\begin{figure}
\centering
\includegraphics[width=0.4\textwidth]{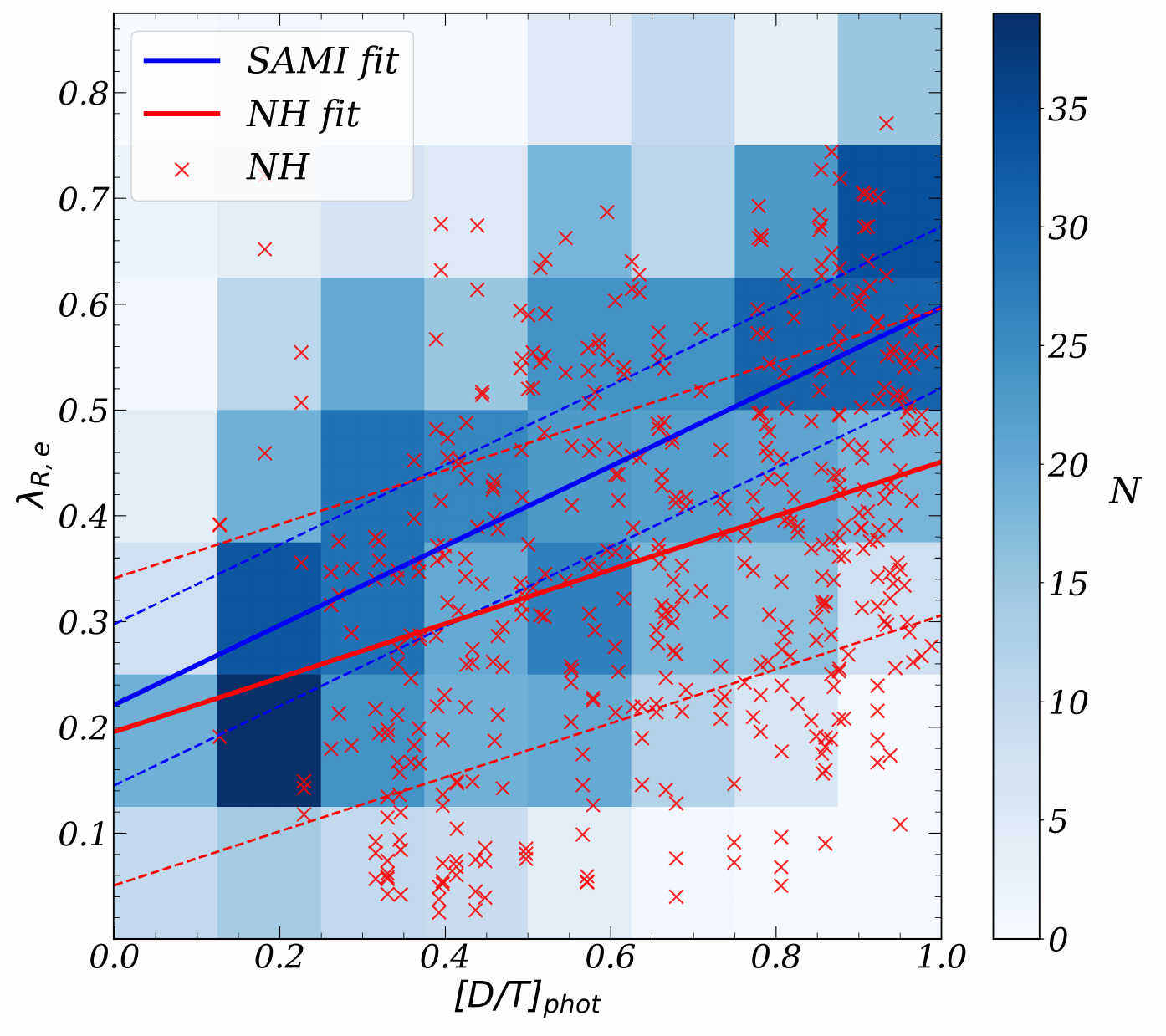}
\caption{
The \spinp\ spin parameter vs. \dttp\ of the \SAMI\ and the \NH\ data. 
The \SAMI\ data are shown in the Hess diagram with a linear fit (blue line) with 1-$\sigma$ errors (dashed lines).
The \NH\ simulation data are also fitted with a linear regression (red lines). 
}
\label{fig:Figure10}
\end{figure}

We show in Figure~\ref{fig:Figure10} the \SAMI\ observed data and \NH\ simulation data in the \spinp\ vs. \dttp\ plane.
The \dttp\, for the \NH\ galaxies has been measured from face-on images, whereas \spinp\ has been measured for the three values of inclinations as mentioned in Section~\ref{sec:spin parameter : method}. 
In addition, for the \dttp\ measurements, two (disk and spheroid) components have been assumed for SAMI, whereas up to three components have been used for \NH, as described in Section~\ref{sec:photometric decomposition}.
As expected based on the discussion in Section~\ref{subsection:photometric result}, there is a very poor correlation between the two parameters in both the observed and simulation data.

\begin{figure}
\centering
\includegraphics[width=0.40\textwidth]{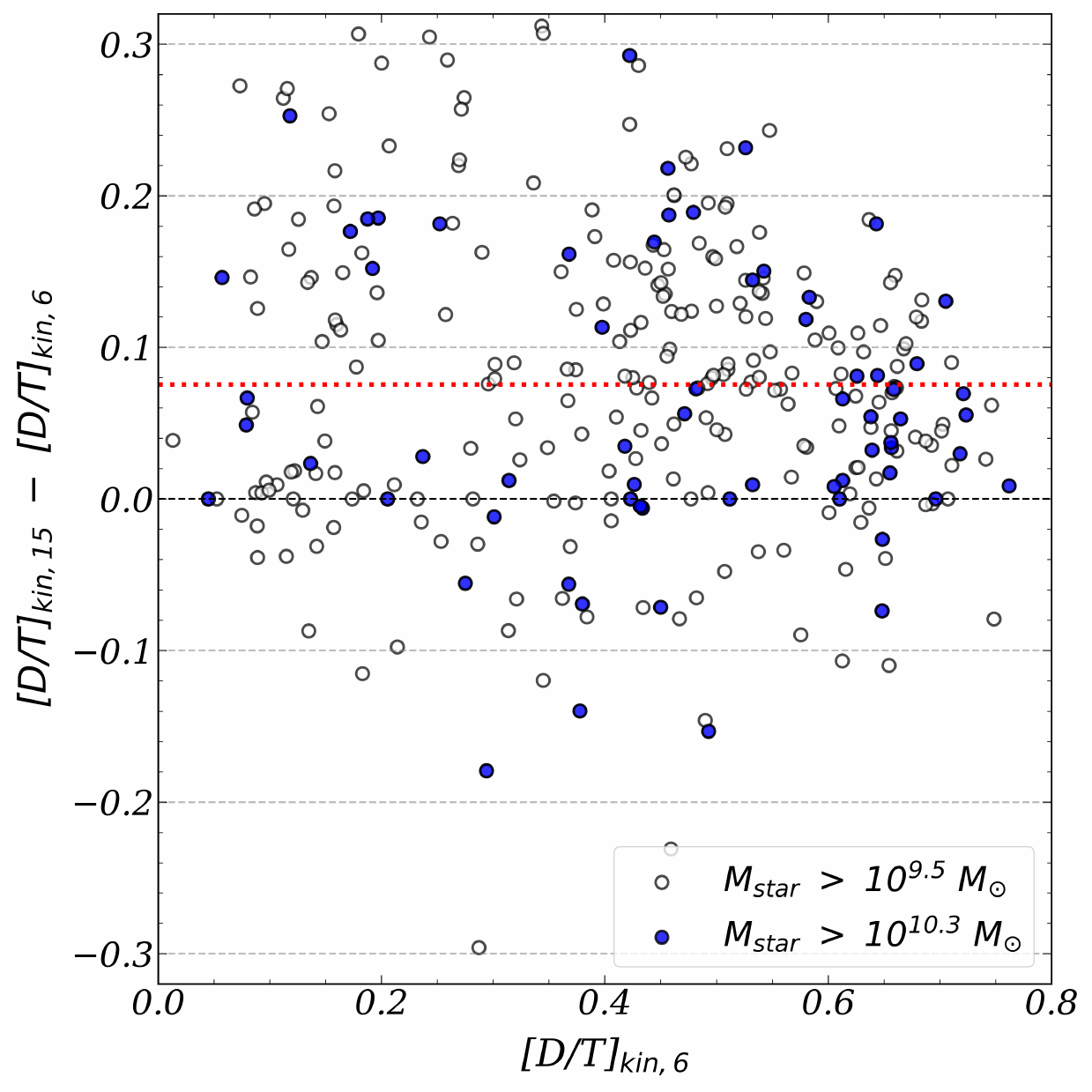}
\caption{
The difference between $[D/T]_{\rm kin,N=6}$ and $[D/T]_{\rm kin,N=15}$.
The red dashed line corresponds to the average offset between the two, which is 0.08. 
}
\label{fig:Figure11}
\end{figure}

It is important but not trivial to understand the cause of the difference between \dttp\ and \dttk.
We can try to assess the impact of technical issues in the measurement of \dttk.
We mentioned earlier that the number of kinematic components in GMM is a free parameter.
The use of a large value can detect even small structures such as tidal streams. 
We used $N_{\rm comp}=6$ in this study to simplify the analysis but have tried other values as large as 15.
Figure~\ref{fig:Figure11} shows the difference in \dttk\ estimates when we use 6 and 15 as component numbers. 
When 15 is used instead of 6, \dttk\ is estimated to be larger by 0.08.
Thus, it is clear that the details of the analysis affect the estimates of \dttk. 
However, the component number accounts for only 20--30\% of the difference between \dttp\ and \dttk.

\begin{figure}
\centering
\includegraphics[width=0.40\textwidth]{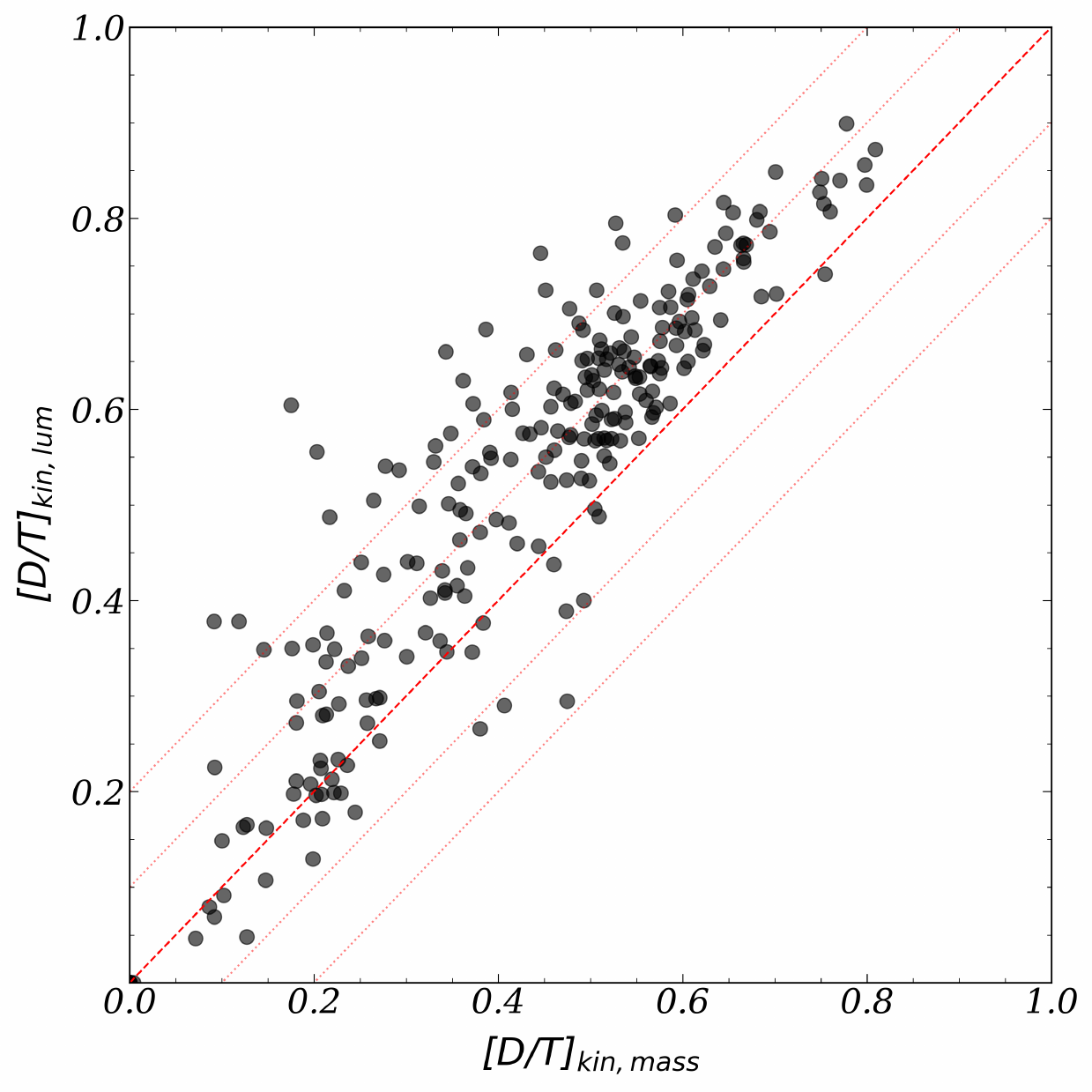}
\caption{
$[D/T]_{\rm kin}$ $r$-band luminosity-weighted versus mass-weighted. The dashed line shows the 1:1 relation, and the four adjacent lines are simple vertical offsets by $\pm 0.1$ and 0.2.
}
\label{fig:Figure12}
\end{figure}

The fact that \dttp\ is luminosity-weighted while \dttk\ is mass-weighted could also cause some difference. 
It is easily conceivable that disk stars being typically younger than other stars cause an increase of \dttp\ in comparison to \dttk.
Figure~\ref{fig:Figure12} indeed shows it.
The luminosity-weighted \dttk\ ratios are higher than the mass-weighted values by roughly 0.1.
However, this effect is so small that when we tried the luminosity-weighted \dttk\ in Figure \ref{fig:Figure9}, the discrepancy between \dttk\ and \dttp\ remained almost the same. 

Another caveat of our kinematic analysis could be that we attempt to detect kinematic components from a smooth and non-distinct distribution in energy and angular momentum. This is obviously not a trivial procedure and thus requires a more elaborate method than just a single cut for example in circularity. We tried to alleviate the issue by using GMM. We however note that in science many or most cases are subject to this issue. Galaxies are often classified as early or late types, gas is often classified as hot or cold gas, and so on, despite that the distribution is not clearly distinct in physical properties, mainly because such a seemingly-simplistic classification is still useful to find some guidelines for understanding complex phenomena.

We also note that our sample of \NH\ galaxies do not have a bar. It is an outstanding issue why high-resolution simulations in particular do not show a bar, as discussed in depth by \citet[][]{Reddish22}. It is unclear how the presence or absence of a bar would affect our results.

Multiple factors affect the measurement of \dttp, too. 
As was the case with GMM, the number of components can be a free parameter in the profile fit, and the use of a different number may affect the measurement, as mentioned in Section \ref{sec:kinematic decomposition}.
Other factors may be as critical as the number of components.
The \sersic\ index is one of them. 
Although we fixed the value for the disk to be 1, we allowed a free value for the other components.
However, it is debatable whether this is the best decision.
The size constraints of the components are also influential.
For instance, are disks always larger than spheroids (in terms of scale length)?
Moreover, the details of the mock-imaging process also cause uncertainties. 
We know that galaxies can follow significantly different dust attenuation laws \citep[][]{FnM90,Calzetti00,Gordon03,Conroy10} depending on the gas fraction and metallicity.
Our \SKIRT\ mock imaging takes the metallicity difference into account; however, the full consideration of attenuation is probably much more complicated than what we have.
It is not trivial to pin down the cause of the departure of \dttp\ from \dttk\ at the moment;
but it would be an important topic for future investigations.

\section{Summary and Conclusion}

This study aimed to inspect the validity of various morphological and kinematic indicators and, if possible, introduce translators between theoretical and observational morphology indicators.
We used the \NH\ simulation, a hydrodynamic cosmological zoom-in simulation with outstanding spatial and mass resolutions. 
We considered galaxies of stellar mass $M_{\rm *} > 10^{9.5}\ \rm M_{\rm \odot}$ sampled from three redshifts ranging from $z=0.17$ to $0.7$. 

We measured the intrinsic kinematic structure using the GMM clustering in a 3-dimensional phase-space (including energy and angular momentum parameters) to group each star particle into one of the six components.
We classified the components with a large mean value of circularity ($\epsilon > 0.5$) as disks.
Then, we defined the kinematic disk-to-total ratio as the mass ratio between the disk components and the total galaxy mass inside \rnt. 

We measured the photometric disk-to-total ratio on the mock images of the galaxies, considering the attenuation from dust using the radiative transfer pipeline SKIRT for a fair comparison of simulated galaxies with observations.
We performed multi-component fitting to radial surface brightness profiles allowing up to four components (nucleus, inner spheroid, disk, and outer spheroid).
The optimal result was selected by comparing the Bayesian information criterion values. 

Moreover, we measured the spectroscopic parameters: \vsig\ and \spinp.
We generated the first and second moment velocity maps using the Voronoi tesselation to ensure an equal $S/N$ ratio.  
We used three values of inclinations to measure the following parameters.

The main results can be summarized as follows.
\begin{itemize}

\item The kinematic disk-to-total ratio reasonably agrees with visual inspection.

\item The spectroscopic parameters exhibited tight correlations with the kinematic disk-to-total ratio.
The \spinp\ spin parameter indicated correlation coefficients in the range of 0.7--0.8, depending on the inclinations.
Similarly-good correlations were found for \vsig.
We provide translators between different indicators.

\item The photometric disk-to-total ratio showed a poor correlation with the kinematic ratio, and a substantial offset (0.2--0.5 in $D/T$) existed.
The photometric decomposition failed to accurately recover the structural composition of galaxies, which seemed more serious for low-mass galaxies that are often classified as pure disks.
While the offsets did not change much, the correlation between the kinematic and photometric disk-to-total ratios became substantially stronger if we removed the low mass galaxies. 
We provide translators between the kinematic and photometric disk-to-total ratios for both cases.

\end{itemize}

Morphology is much more than just a first impression.
\citet[][]{Hubble1926} correctly noticed that it contains important information for the nature of galaxies. 
Since then abundant information regarding the relationship between the apparent morphology and true properties of galaxies has been obtained.
Galaxies are thought to be composed mainly of a rotation-dominant component and dispersion-dominant component.
However, it is almost certain that reality is much more complex. 
While observational astronomy significantly contributed to galaxy research in the last century, we have just succeeded in making arguably-realistic models of galaxies in cosmological large-volume numerical simulations. 
The question is how realistic they are.
Observations and simulations use different languages, and translators are required.
In this study, we attempted to find such a translator.
We found that spectroscopic indicators, such as \vsig\ and \spinp\, closely traced the true kinematic structure of galaxies.
In contrast, the photometric profile fits failed to recover it accurately, especially for small galaxies.
We were able to find mappings that could be useful for galaxies with good observational quality.
We hope that this translator will be useful when simulations are tested against observations.
There are a few open questions: for example, the issue of pure disks and the implication of multiple components in the GMM analysis, which require further investigation.

\begin{acknowledgments}
\section*{Acknowledgements}

This work was granted access to the HPC resources of CINES under the allocations  c2016047637, A0020407637 and A0070402192 by Genci, KSC-2017-G2-0003 by KISTI, and as a “Grand Challenge” project granted by GENCI on the AMD Rome extension of the Joliot Curie supercomputer at TGCC. 
This research is part of the Spin(e)  ANR-13-BS05-0005 (\href{http://cosmicorigin.org}{http://cosmicorigin.org}), Segal ANR-19-CE31-0017 (\href{http://secular-evolution.org}{http://secular-evolution.org}) and Infinity-UK projects.
This work has made use of the Horizon cluster on which the simulation was post-processed, hosted by the Institut d'Astrophysique de Paris. We warmly thank S.~Rouberol for  running it smoothly.
The large data transfer was supported by KREONET which is managed and operated by KISTI. 
The radiative transfer pipeline, SKIRT, played a pivotal role in this study. We sincerely thank its creators: Maarten Baes and Peter Camps. 
S.K.Y. acknowledges support from the Korean National Research Foundation (NRF-2020R1A2C3003769). 
TK was supported in part by the National Research Foundation of Korea (NRF-2020R1C1C1007079).
This study was in part funded by the NRF-2022R1A6A1A03053472 grant and the BK21Plus program. 
Parts of this research were supported by the Australian Research Council Centre of Excellence for All Sky Astrophysics in 3 Dimensions (ASTRO 3D), through project number CE170100013.
\\
\vspace{-0.5cm}
\end{acknowledgments}



\appendix
\restartappendixnumbering

\section{Phase-space distribution}
\label{sec appendix_A}

\begin{figure}[h]
\centering
\includegraphics[width=0.45\textwidth]{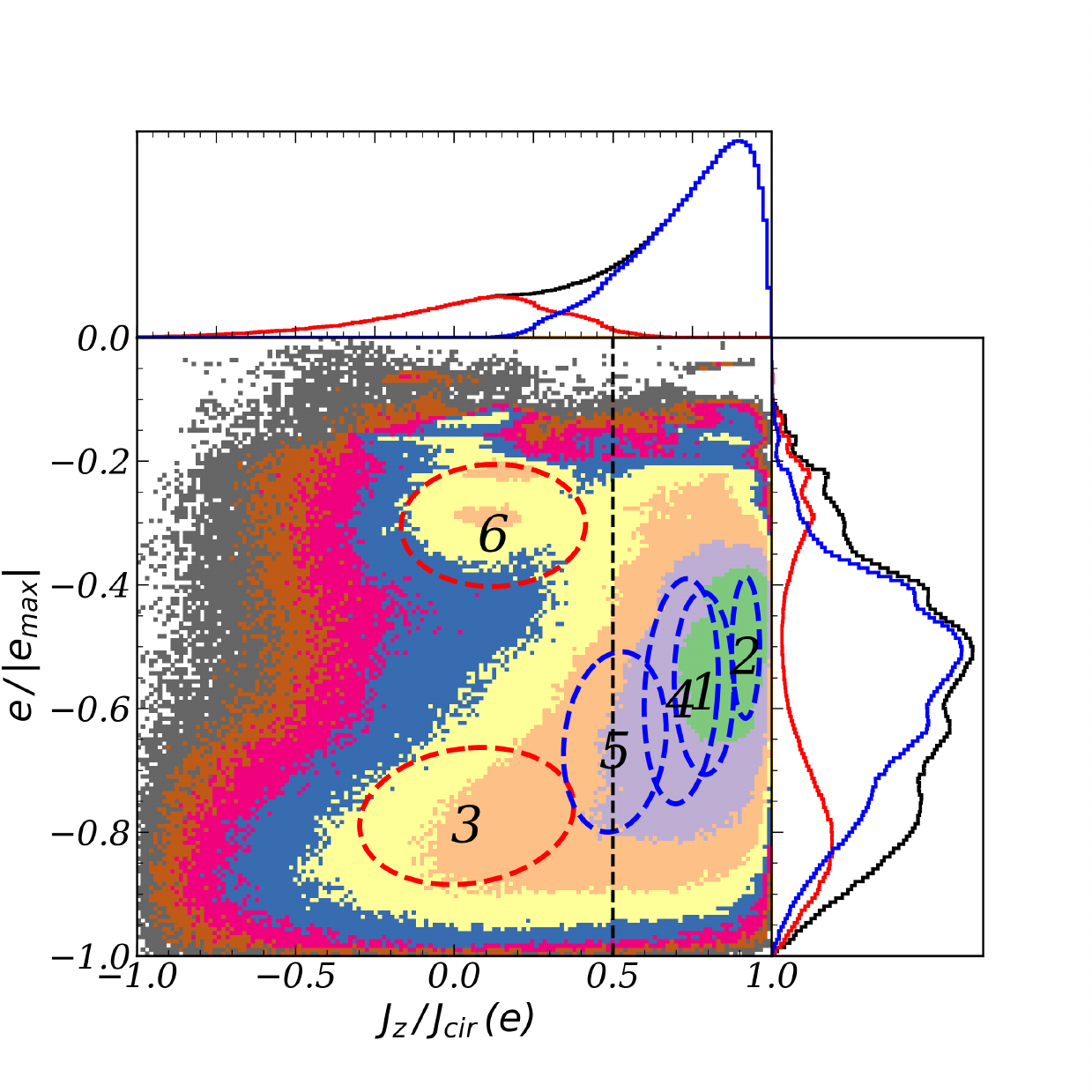}
\caption{
The phase-space distribution of a disk galaxy without a nucleus component in the same format as Figure~\ref{fig:Figure2}.
}
\label{fig:FigureA1}
\end{figure}

\begin{figure*}[h]
\centering
\includegraphics[width=0.85\textwidth]{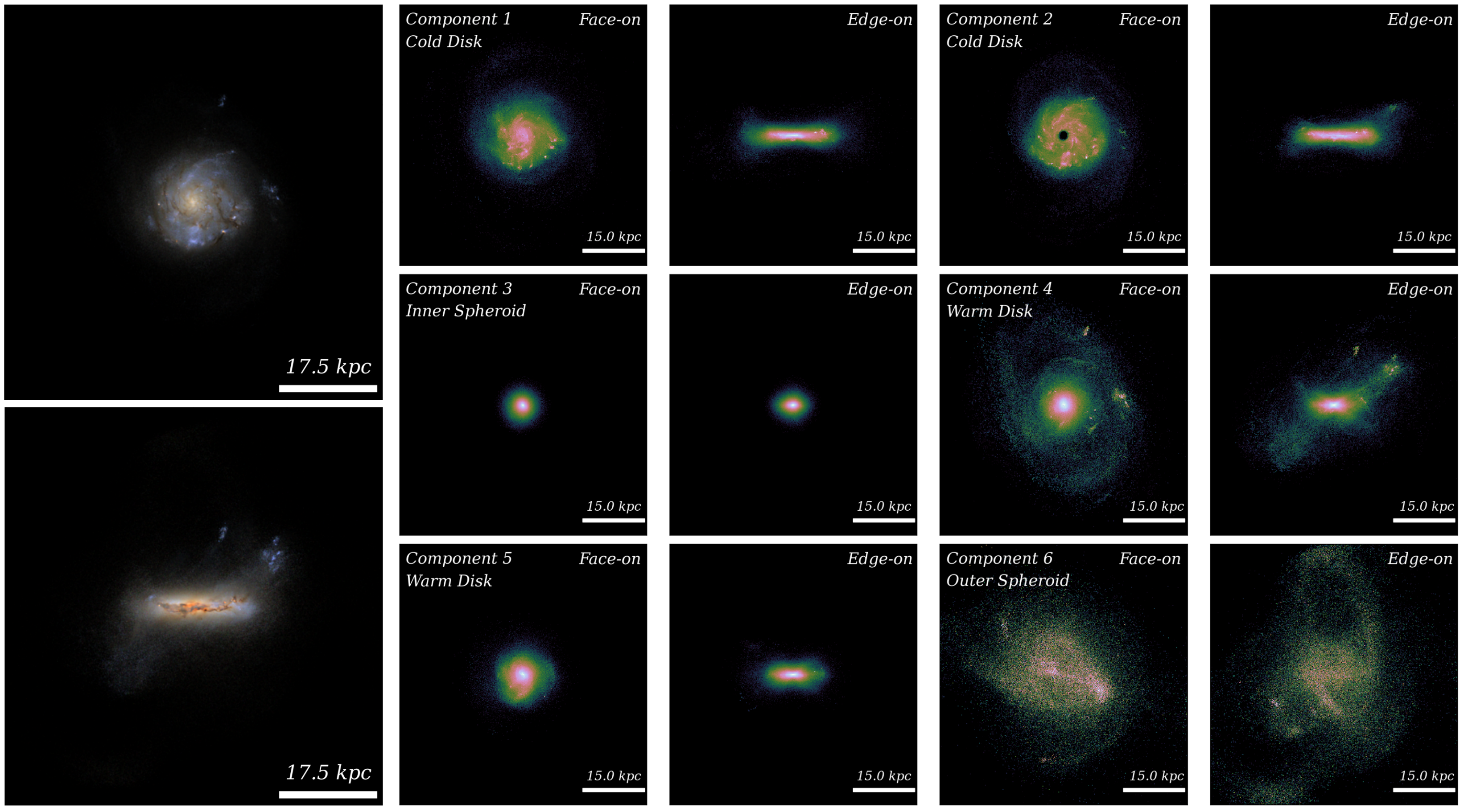}
\caption{
The face-on and the edge-on $r$-band flux density map of an example galaxy and its GMM components detected in Figure~\ref{fig:FigureA1}, in a similar format to Figure~\ref{fig:Figure3}.
}
\label{fig:FigureA2}
\end{figure*}

We select another sample galaxy from the NewHorizon simulation, this time without a nucleus component.
We present in Figure~\ref{fig:FigureA1} the phase-space distribution and the detected GMM components in the same format as in Figure~\ref{fig:Figure2}.
As we mentioned in section~\ref{sec:kinematic decomposition}, the phase-space distribution of the detected components is different from that shown in Figure~\ref{fig:Figure2}.
We also present in Figure~\ref{fig:FigureA2} the spatial distribution of corresponding GMM components in a similar format to Figure~\ref{fig:Figure3}.
Note that the GMM analysis detected two warm disks but no nucleus in this galaxy. 
In addition, tidal features are clearly visible in the outer spheroid.

\section{$v/\sigma$-circularity relation}
\label{sec appendix_B}

\begin{figure*}[h]
\centering
\includegraphics[width=0.5\textwidth]{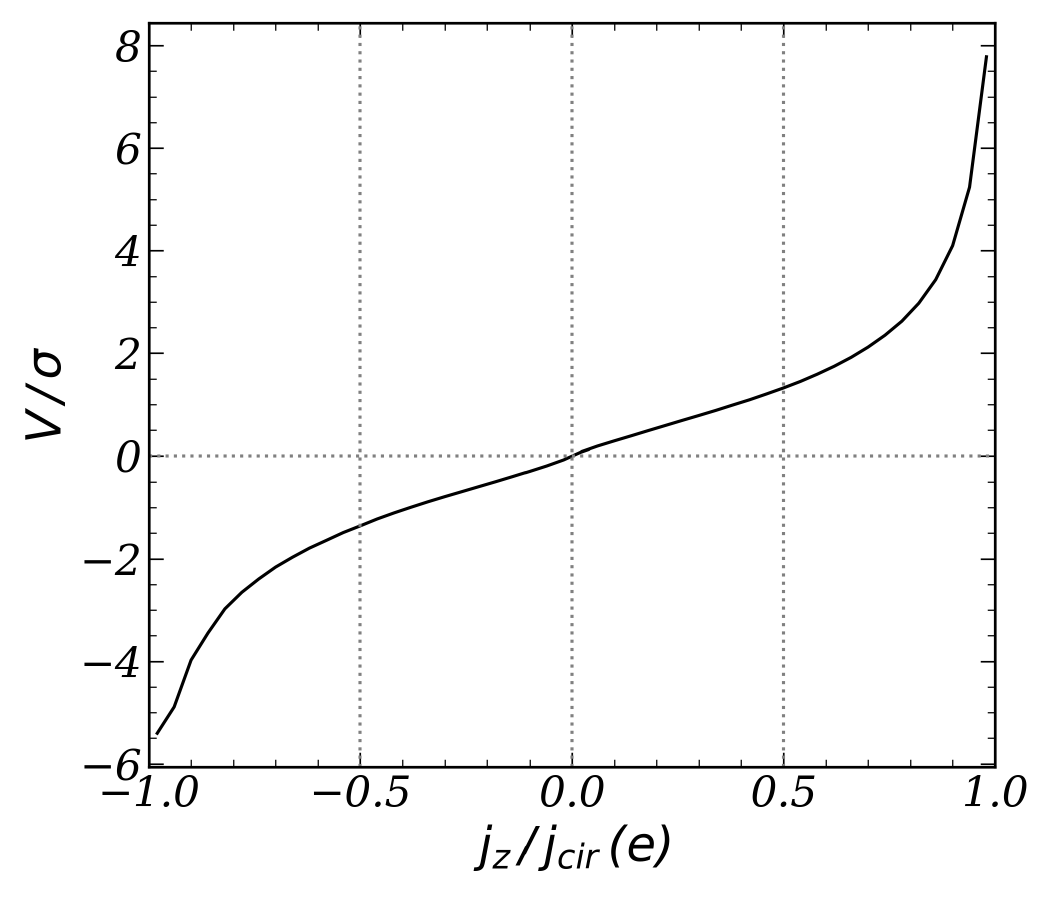}
\caption{
$v/\sigma$-circularity relation based on the galaxy in Figure~\ref{fig:Figure2}. 
}
\label{fig:FigureB3}
\end{figure*}

We provide the $v/\sigma$-circularity relation based on one NH galaxy in Figure~\ref{fig:FigureB3}. $\epsilon=0.5$ slightly overshoots $v/\sigma=1$. In this galaxy, the disc is dominant and likely has a tail below $\epsilon=0.5$. On the contrary, if a galaxy is dispersion dominant, the tails of the dispersion components would overshoot $\epsilon=0.5$. So, using a single cut (of circularly or $v/\sigma$) is not effective. We use the machine learning procedure, GMM, that takes into account the highest density peaks and their tail distributions in our investigation to evade such complexities.

\end{document}